\documentclass{jov}

\usepackage[utf8]{inputenc}
\usepackage[T1]{fontenc}
\usepackage{lmodern}
\usepackage{amsmath}
\usepackage{amsfonts}
\usepackage{amssymb}
\usepackage{tabularx}
\usepackage{natbib}
\usepackage{gensymb}
\usepackage{units}
\usepackage{eurosym}
\usepackage[english]{babel}
\usepackage[normalem]{ulem}

\PassOptionsToPackage{pdftex}{graphicx}
\usepackage{graphicx} 
\usepackage{xcolor}
\definecolor{orange}{rgb}{.5,.35,0}

\begin{document}

\title{Disentangling bottom-up vs.~top-down and low-level \\[0.5ex] vs.~high-level influences on eye movements over time}

\abstract{Bottom-up and top-down, as well as low-level and high-level factors influence where we fixate when viewing natural scenes. However, the importance of each of these factors and how they interact remains a matter of debate. Here, we disentangle these factors by analysing their influence over time. For this purpose we develop a saliency model which is based on the internal representation of a recent early spatial vision model to measure the low-level bottom-up factor. To measure the influence of high-level bottom-up features, we use a recent DNN-based saliency model. To account for top-down influences, we evaluate the models on two large datasets with different tasks: first, a memorisation task and, second, a search task. Our results lend support to a separation of visual scene exploration into three phases: The first saccade, an initial guided exploration characterised by a gradual broadening of the fixation density, and an steady state which is reached after roughly 10 fixations. Saccade target selection during the initial exploration and in the steady state are related to similar areas of interest, which are better predicted when including high-level features. In the search dataset, fixation locations are determined predominantly by top-down processes. In contrast, the first fixation follows a different fixation density and contains a strong central fixation bias. Nonetheless, first fixations are guided strongly by image properties and as early as 200\,ms after image onset, fixations are better predicted by high-level information. We conclude that any low-level bottom-up factors are mainly limited to the generation of the first saccade. All saccades are better explained when high-level features are considered, and later this high-level bottom-up control can be overruled by top-down influences.}

\author{Schütt*}{Heiko H.}
 {Neural Information Processing Group, Universität Tübingen, Tübingen, Germany}
 {and Experimental and Biological Psychology, University of Potsdam, Potsdam, Germany}
 {http://}{Heiko.schuett@uni-tuebingen.de}
\author{Rothkegel*}{Lars O. M.}
 {Experimental and Biological Psychology}
 {University of Potsdam, Potsdam, Germany}
 {http://}{lars.rothkegel@uni-potsdam.de}
\author{Trukenbrod}{Hans A.}
 {Experimental and Biological Psychology}
 {University of Potsdam, Potsdam, Germany}
 {http://}{hans.trukenbrod@uni-potsdam.de}
\author{Engbert}{Ralf}
 {Experimental and Biological Psychology}
 {University of Potsdam, Potsdam, Germany}
 {http://}{ralf.engbert@uni-potsdam.de}
\author{Wichmann}{Felix A.}
 {Neural Information Processing Group}
 {Universität Tübingen, Tübingen, Germany}
 {http://}{felix.wichmann@uni-tuebingen.de}

\keywords{saliency, fixations, natural scenes, visual search, eye movements}

\maketitle

{\small \noindent * Heiko H. Schütt and Lars O. M. Rothkegel contributed equally to this work.}

\newpage

\section{Introduction}
The guidance of eye movements in natural environments is extremely important for our perception of the world surrounding us. Visual perception deteriorates quickly away from the gaze position such that many tasks are hard or impossible to perform without looking at the objects of interest \citep[reviewed by][Section 6; see also: \citealt{land1999}]{strasburger2011}. Thus, the selection of fixation locations is of great interest for vision researchers and many theories were developed to explain the selection of fixation locations.

Classically, factors determining eye movements of human observers are divided into bottom-up and top-down influences \citep{hallett1978, tatler2008}. Bottom-up influences refer to stimulus parts which attract fixations independent of the internal state of an observer. The existence of bottom-up guidance of eye movements was originally postulated because some stimuli like flashing lights attract subjects' gaze under well controlled laboratory conditions, even in tasks when subjects were explicitly asked not to look at the stimulus as, for example, in the anti-saccade task \citep{hallett1978, klein2001, mokler1999, munoz2004}. How important bottom-up effects are under more natural conditions and especially for static stimuli remains a matter of debate.
Top-down influences on the other hand refer to cognitive influences on the chosen fixation locations, based on the current aims of an observer varying, for example, with task demands and memory \citep{land1999,henderson2007}. The main argument for the involvement of top-down control comes from task effects on fixation locations \citep{einhauser2008a, underwood2006, henderson2007,yarbus1967}.
More recently, systematic tendencies were introduced as a third category \citep{tatler2008}, which encompass regularities of the oculomotor system across different instances and manipulations like the preference to fixate near the image center \citep{tatler2007}, the preference for some saccade directions \citep{foulsham2008a}, and the dependencies between successive saccades \citep{tatler2008, wilming2013, rothkegel2016}. While all three aspects seem to contribute to eye movement control, the debate, how these aspects are combined and how important the different aspects are, continues till today \citep{hallett1978, harel2006, tatler2009, kienzle2009, einhauser2008a, foulsham2008b, tatler2011, borji2013, stoll2015, schomaker2017}.

Orthogonal to the top-down vs.~bottom-up distinction, models of eye movement control can also be categorised by the features they employ. Here low-level models refer to simple features, which are extracted early in the visual hierarchy like local colour, brightness or contrast \citep{itti2001, treisman1980}. High-level models on the other hand refer to features thought to be extracted in higher cortical areas encodig more complex information like the position and identity of objects \citep{einhauser2008b} or the scene category and context \citep{torralba2006}.

In the debate on what factors govern eye movements, the two sides typically argued for are, on the one hand, bottom-up control based on low-level features \citep{itti2001, kienzle2009} and, on the other hand, top-down control based on high-level features \citep{castelhano2009, yarbus1967}. This links the question of feature complexity to the question how much internal goals control our eye movements. These questions are orthogonal, however, and the less usual positions may be sensible as well. Bottom-up control may encompass not only low-level features like contrast, colour or edges \citep{treisman1980,itti1998,itti2001}, but also high-level properties of the explored scene like object locations \citep{einhauser2008b}, faces \citep{judd2009,kummerer2016} or even locations which are interesting or unexpected in a scene category \citep{henderson1999b, torralba2006}. This position is implicitly embraced by most modern (computer vision) models for the prediction of fixation locations in images, which almost all use high-level features computed from the image \citep{judd2009,mit-saliency-benchmark,kummerer2016}. Similarly, top-down control may not only act on high-level features but can also act on low-level features like contrast, orientations or colour. Such influences are especially important in models of visual attention \citep{tsotsos1995, muller2006} and visual search \citep{wolfe1994}, which often postulate top-down control over low-level features that guide attention and eye movements or top-down influences on the processing of low-level features \citep{tsotsos1995}.

In addition, inconsistent terminology has added to the confusion. One especially unclear term in this context is saliency, which originally stems from attention research. Saliency referred to conspicuous locations which stood out from the remaining display and attracted attention \citep{koch1985}. As the first computable models for the prediction of fixation locations in images were based on these ideas, saliency was soon associated with these models and became synonymous with bottom-up low-level control of eye movements \citep{itti1998,itti2001}. As it became clear that the prediction of fixation locations benefits from high-level features, they were added to the models, but the models were still referred to as saliency models \citep{judd2009,mit-saliency-benchmark,borji2013}. As a result, saliency in computer vision now refers to any bottom-up image-based prediction of which locations are likely to be fixated by human observers. To avoid confusion associated with the term saliency, we will use the term "saliency model" in the remainder, which refers to any bottom-up model that predicts fixation locations based on an image, independent of feature complexity.

In debates about eye movement control, the temporal evolution of the fixation density over a trial has been largely ignored \citep[see][for notable exceptions]{anderson2015, anderson2016, stoll2015}. However, the temporal evolution might be informative to understand the interplay of the different factors. To fill this gap with this article, we analyse the temporal evolution of the fixation density and disentangle the contribution of the different factors.

To allow stable estimates of the fixation density at different ordinal fixation numbers (fixation \#s), we employ two large eye movement data sets, which contain exceptionally many fixations per image. The two datasets were collected using different tasks, which allows us to analyse, whether our conclusions hold under different top-down control conditions.
The first data set stems from a standard scene viewing experiment in which participants were asked to explore a scene for a subsequent memory test. Scene viewing has been suggested to minimise top-down control \citep{itti2001}, although subjects could still employ top-down control and different subjects could even choose different top-down control strategies \citep{tatler2011}. For the second data set, subjects searched for artificial targets in natural scenes \citep{rothkegel2018}. Since the search target was known in advance, participants had a clear aim and motivation to exploit low-level features in a top-down fashion in the second task.

Using these two data sets we systematically investigate how fixation locations are controlled by low- and high-level features under the different top-down control demands. To quantify the contribution of low- and high-level features, we compare the performance of different computational models using a recently proposed likelihood based technique \citep{kummerer2015,schutt2017}. This method avoids ambiguities of using typical saliency model evaluation criteria and provides a unified metric for all models. 

To measure the influence of low-level features, we choose three classical low-level models \citep[][]{itti1998,harel2006,kienzle2009}. However, the features used by these low-level models were only informally linked to the low-level features used in models of perception. To remove this ambiguity of interpretation we additionally present a new saliency model, which is based on the representation produced by a successful model of early spatial vision \citep{schutt2017b}. 

To measure the influence of high-level features, earlier approaches made predictions based on manually annotated object locations \citep{einhauser2008b, torralba2006, stoll2015}, experimentally varied low-level features \citep{acik2009, stoll2015, anderson2015} or chose specific examples for which low-level and high-level features make opposing predictions \citep{vincent2009}. But these classical approaches do not easily make predictions for new images. Fortunately, the idea of object-based saliency models was recently unified  with low-level factors due to the advent of deep neural network models \citep[DNNs; see][for an overview]{kriegeskorte2015}.  DNNs contain activation maps which effectively encode what kind of object can be found where in an image. Like simple low-level features, these object-based features can be used to predict fixation locations \citep{kummerer2016, pan2017, huang2015, kruthiventi2015}. Saliency models based on this principle currently perform best on the fixation density prediction benchmarks \citep{mit-saliency-benchmark}. Thus, these DNN-based saliency models provide a better and more convenient quantification than earlier approaches about which information can be predicted by high-level features. As a representative we use the currently best performing of these models, DeepGaze II \citep{kummerer2016}.

\section{Methods}

\subsection{Stimulus presentation}
Sets of 90 (scene viewing experiment) and 25 (search experiment) images were presented on a 20-inch CRT monitor (Mitsubishi Diamond Pro 2070; frame rate 120~HZ, resolution 1280$\times$1024 pixels; Mitsubishi Electric Corporation, Tokyo, Japan). All stimuli had a size of 1200$\times$960 pixels. For the presentation during the experiment, images were displayed in the center of the screen with grey borders extending 32 pixels to the top/bottom and 40 pixels to the left/right of the image. The images covered $31.1^\circ$~of visual angle in the horizontal and $24.9^\circ$~in the vertical dimension. 

\subsection{Measurement of eye movements}
Participants were instructed to position their heads on a chin rest in front of a computer screen at a viewing distance of 70~cm. Eye movements were recorded binocularly using an Eyelink 1000 video-based-eyetracker (SR-Research, Osgoode/ON, Canada) with a sampling rate of 1000~Hz. 

For saccade detection we applied a velocity-based algorithm \citep{engbert2003,engbert2006}. The algorithm marks an event as a saccade if it has a minimum amplitude of 0.5$^\circ$~and exceeds the average velocity during a trial by 6 median-based standard deviations for at least 6 data samples (6 ms). The epoch between two subsequent saccades is defined as a fixation. All fixations with a duration of less than 50~ms were removed for further analysis since these are most probably glissades, i.e., part of the saccade \citep{nystrom2010}. 
The number of fixations for further analyses was $312\,267$ in the scene viewing experiment and $176\,828$ in the search experiment.

For calibration we performed a 9 point calibration in the beginning of each session of the scene viewing experiment and of each block of the search experiment and recalibrated every ten trials or whenever the fixation check at the beginning of a trial failed. 

\begin{figure}
\unitlength1mm
\begin{picture}(160,75)
\put(0,10){Scene Viewing Dataset}
\put(0,15){\includegraphics[width=7.5cm]{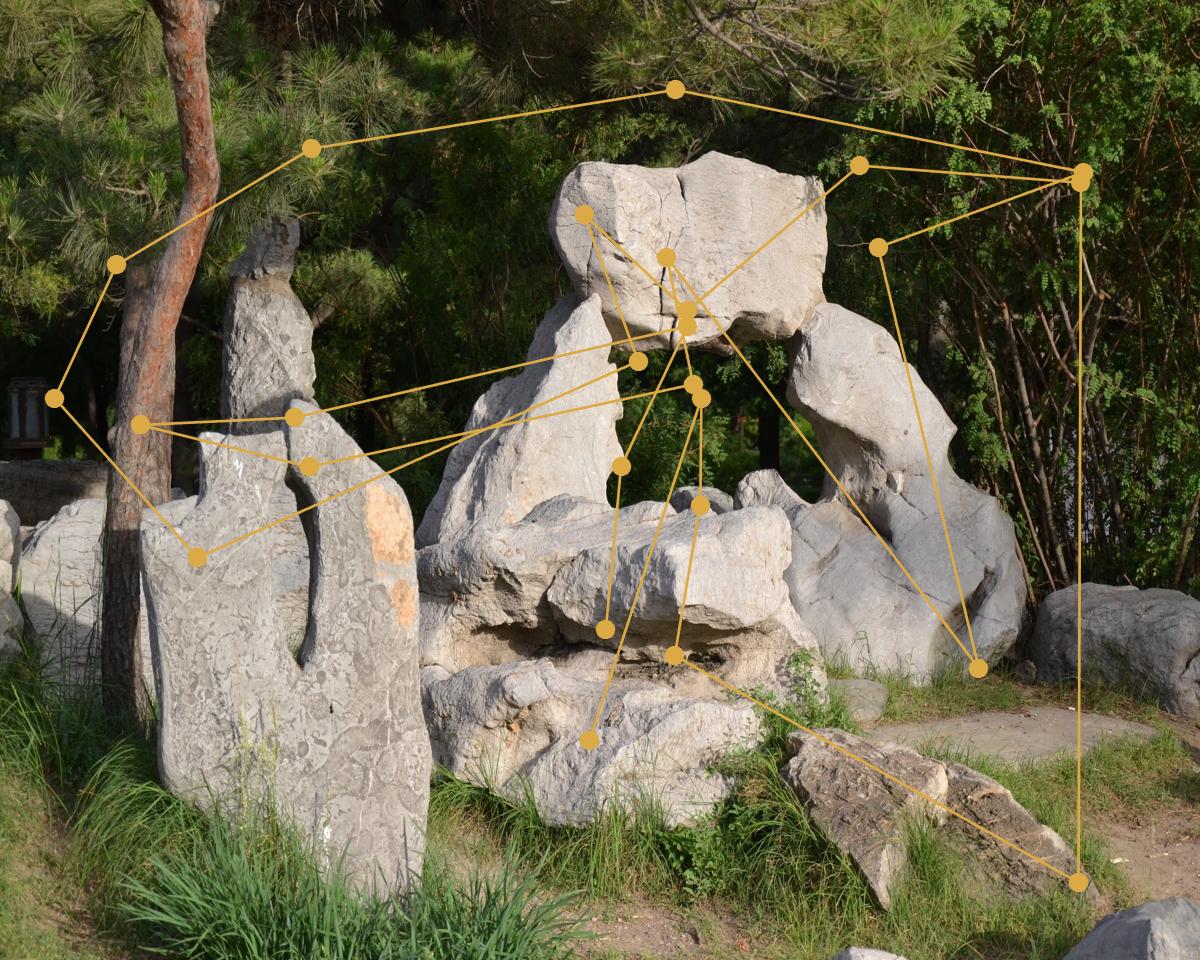}}
\put(85,0){\includegraphics[height=7.5cm]{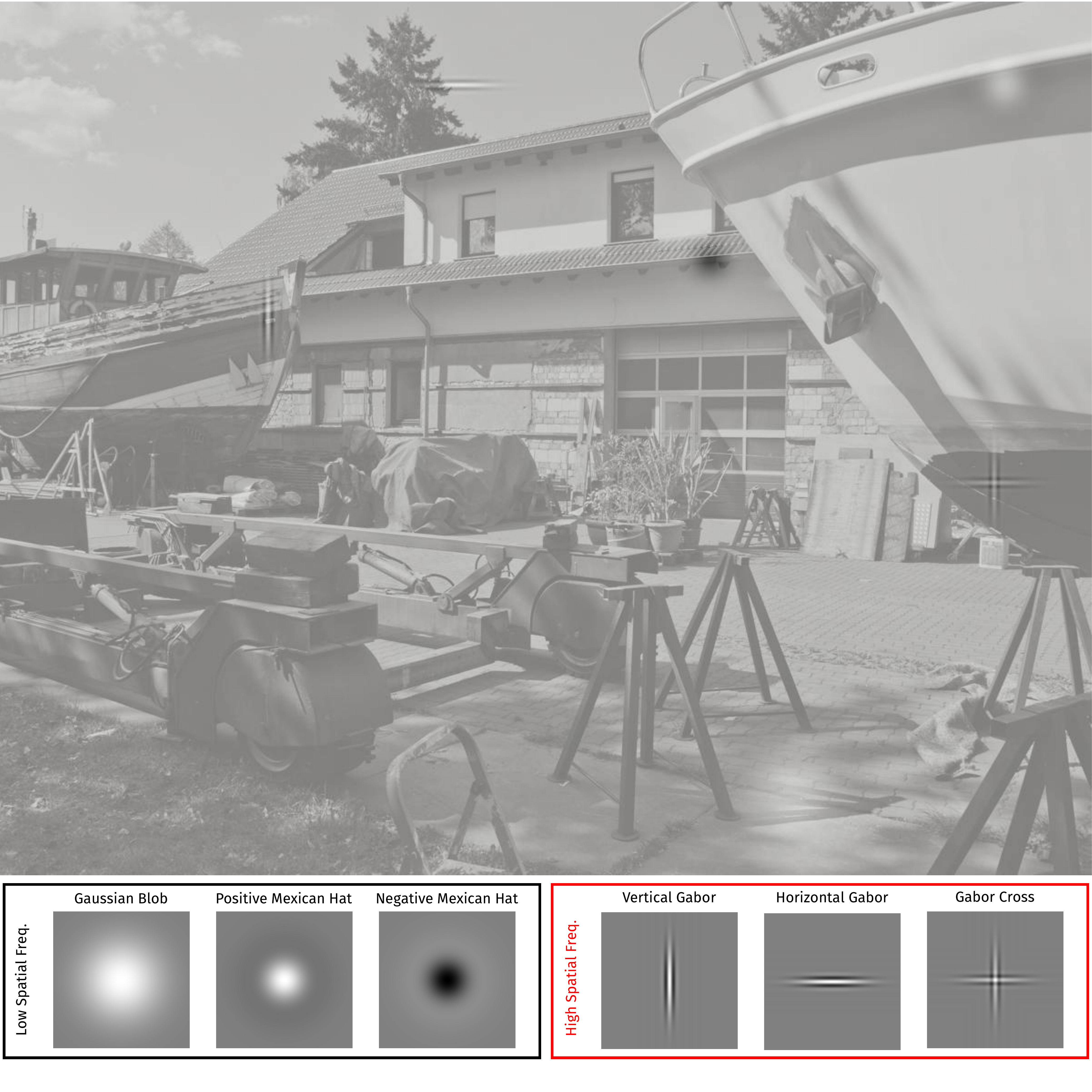}}
\end{picture}
\caption{Overview over datasets. Left: Image from scene viewing dataset with exemplary scanpath. We recorded eye movements of 105 subjects on the same 90 images with slightly varying viewing conditions asking them to remember which images they had seen for a subsequent test.  Right: visual search task. Here we recorded eye movements of 10 subjects searching for the 6 targets displayed below the image for eight sessions each. In the experiment each image contained only one target and subjects usually knew which one. Additionally we increased the size and contrast of the targets for this illustration image to compensate for the smaller size of the image. The right panel is reused with permission from our article on the search dataset \citep{rothkegel2018}.\label{fig:Datasets}}
\end{figure}

\subsection{Scene Viewing Dataset}
In our scene viewing experiment we showed 90 images to 105 participants in three groups with slightly varying viewing conditions, asking them to remember the presented images for a subsequent memory test. 
\subsubsection{Participants}
For this study 105 students of the University of Potsdam with normal or corrected to normal vision were recruited. On average participants were 23.3 years old and 89 participants were female. Participants received credit points or a monetary compensation of 16\euro~for their participation. The work was carried out in accordance with the Declaration of Helsinki. Informed consent was obtained for experimentation by all participants.

\subsubsection{Stimuli}
As stimuli we used 90 photographs taken by the authors, which did not contain text or humans that stand out prominently. Furthermore, images were selected as 6 subsets of 15 images each: The first contained photographs of texture-like patterns, the other 5 contained typical holiday photographs with the prominent structure either at the top, left, bottom, right or center. The full set of images is available online with the dataset (Fig.~\ref{fig:Datasets}, left panel). 

For presentation in greyscale, we measured the luminance output $[\frac{cd}{m^2}]$ of each gun separately and for the sum of all three guns at every value from 0 to 255. To convert a stimulus into greyscale, we summed the luminance output for the RGB values and chose the grey value with the most similar luminance.

\subsubsection{Procedure}
Eye movements for our scene viewing experiment were collected in two sessions. In each session 60 images were presented and participants were instructed to memorise them for a subsequent test, to report which images they had seen. In the first session all images were new. In the second session we repeated 30 images from the first session and showed the remaining 30 new images. The 30 repeated images were the same for each observer. For this article we used all fixations from both sessions, ignoring whether the subject had seen the image before and which group the subject belonged to, to maximise the amount of data. 
Trials began with a black fixation cross presented on a grey background. After successful binocular fixation in a square with a side length of $2.2^\circ$ the stimulus appeared and subjects had 10 seconds to explore the image.
In the memory test participants had to indicate for 120 images if they had seen it before. Half the images were the ones they saw in the experiment, the other half were chosen randomly from another pool of 90 images we chose according to the same criteria as the images used for the first set of images. 

The three cohorts of subjects differed in the placement of the fixation cross and whether the images were shown in colour or in greyscale:
\begin{itemize} 
\item For the first 35 subjects we presented the images in greyscale and placed the start position randomly within a doughnut-shape around the center of the screen and stimulus, with an inner radius of $100px =2.6^\circ$ and an outer radius of $300px=7.8^\circ$.
\item For the second group of 35 subjects the images were also presented in greyscale, but the start position was chosen randomly from only 5 positions: The image center and 20\% of the monitor size (256/205 pixels, $5.68/4.55$ degree of visual angle) away from the border of the monitor at the top, left, bottom and right, centrally in the other dimension.
\item For the final group of 35 subjects the images were shown in colour and the starting position was as for the second group. 
\end{itemize}

\subsection{Natural image search}
In our visual search task, participants were asked to look for six targets embedded into natural scenes (Fig.~\ref{fig:Datasets}). The data set has been used for a different purpose in another publication \citep{rothkegel2018}.

\subsubsection{Participants}
We recorded eye movements from 10 human participants (4 female) with normal or corrected-to-normal vision in 8 separate sessions on different days. 6 participants were students from a nearby high school (age 17 to 18) and 4 were students at the University of Potsdam (age 22 to 26). 

\subsubsection{Stimuli} 
As natural image backgrounds we chose 25 images taken by the authors and an additional member of the Potsdam lab in the area surrounding Potsdam. The images contained neither faces nor text. 

As targets we designed 6 different low-level targets with different orientation and spatial frequency content (Fig.\ref{fig:Datasets}, right panel). To embed the targets into natural images we first converted each image to luminance values based on a power function fitted to the measured luminance response of the monitor. We then combined this luminance image $I_L$ with the target $T$ with a luminance amplitude $\alpha L_{max}$ fixed relative to the maximum luminance displayable on the monitor $L_{max}$ as follows:
\begin{equation}
I_{fin} = \alpha L_{max} + (1-2\alpha)I_L +\alpha L_{max} T,
\end{equation}
i.e., we rescaled the image to the range $[\alpha ,(1-\alpha)]L_{max}$ and then added the target with a luminance amplitude of $\alpha L_{max}$, such that the final image $I_{fin}$ never left the displayable range. After a pilot experiment we fixed $\alpha$ to 0.15. Thus, contrast was reduced to $70\%$. We then converted the image $I_{fin}$ back to $[0,255]$ greyscale values by inverting the fitted power function.

\subsubsection{Procedure}
Participants were instructed to search for one of 6 targets for the upcoming block of 25 images. To do so, the target was presented on a 26th demonstration image, marked by a red square. Each session consisted of 6 blocks of 25 images for each of the 6 different targets. The 25 images within a block were always the same presented in a new random order. 

Trials began with a black fixation cross presented on grey background at a random position within the image borders. After successful fixation, the image was presented with the fixation cross still present for 125~ms. This was done to assure a prolonged first fixation to reduce the central fixation tendency of the initial saccadic response \citep{tatler2007,rothkegel2017}. After removal of the fixation cross, participants were allowed to search the image for the previously defined target for 10~s. Participants were instructed to press the space bar to stop the trial once a target was found. In $\sim80\%$ of trials the target was present. 

At the end of each session participants could earn a bonus of up to 5\euro~additional to a fixed 10\euro~reimbursement, depending on the number of points collected divided by number of possible points. If participants correctly identified a target, they earned 1 point. If participants pressed the bar although no target was present, one point was subtracted.

\subsection{Kernel density estimation of fixation densities}
To estimate empirical fixation densities, we used kernel density estimation as implemented in the R package \emph{spatstat} (version 1.51-0). Kernel density estimation requires the choice of a bandwidth for the kernel. The optimal choice for this parameter depends on the shape of the density and on the number of observations available. Thus it cannot be chosen optimally a priorix, but needs to be chosen adaptively for each condition.

To set the bandwidth for our kernel density estimates we used leave one subject out cross-validation, i.e., for each subject we evaluated the likelihood of their data under a kernel density estimate based on the data from all other subjects. For the image dependent density estimates we repeated this procedure with bandwidths ranging from 0.5 to 2.0 degrees of visual angle (dva) in steps of 0.1 dva. We report the results with the best bandwidth chosen for each fixation \# separately. For the image independent prediction---i.e., the central fixation bias---we used the same procedure with bandwidths from 0.2 to 2.2 dva as these estimates are based on more data and chose a single bandwidth over all images. 

To implement this procedure, we calculated the cross-validated log-likelihood for each fixation using each possible bandwidth. This calculation results in a 4 dimensional array with dimensions for the fixation \#, the subject, the image and the chosen bandwidth. We then averaged over subjects and images and report the highest value for each fixation \#.

For our analysis over time we calculated two estimates of the fixation density as upper bounds for the predictability of fixation locations from a static map. For the first we simply took the cross-validated kernel density estimate based only on the fixations with the same fixation \# (labelled "Empirical Density (each \#). Fixations typically become fewer later in the trial, as fixation durations and thus the number of fixations within a 10 second trial varies. Furthermore, we observe that fixations are more dispersed later in the trial, such that more fixations are required to estimate the fixation density accurately. Thus, our first estimate declines rapidly over time. To counteract this, we computed a second estimate which uses all fixations on the image from the second to the last fixation to predict the density (labelled "Empirical Density (all \#)"). This estimate can use more data and performs well, because the fixation density converges towards the end of the trial (see Fig.\ref{fig:CorpusTime}).

The likelihood of the kernel density estimates always depended smoothly on the bandwidth and showed a single peak such that the bandwidth could be chosen reliably. Furthermore, the chosen bandwidths behaved as expected, such that estimates of the same type for later fixation densities were made with larger bandwidths. For the scene viewing dataset the bandwidth varied from 0.5 to 0.7 dva for the empirical density and from 1.1 to 1.8 dva for the central fixation bias. For the search experiment the bandwidth for the empirical density varied from 0.7 to 1.1 dva and the bandwidth for the central fixation bias varied from 1.3 to 2.0 dva. Suboptimal choices of the bandwidth could lead to arbitrarily bad performance. Within the range we observed differences between bandwidths of less than $0.4\frac{bit}{fix}$ for the scene viewing experiment and less than $0.2\frac{bit}{fix}$ for the search experiment. Thus, not choosing the bandwidth optimally we would noticeably underestimate the absolute performance of the two estimates, but the results would not vary qualitatively.

\subsection{Comparing fixation densities}

To compare two fixation densities $p_1,p_2$ we computed a kernel density estimate $\hat{p}_1$ for one of the fixation densities $p_1$ and evaluated the log-likelihood of the fixations $f^{(i)}_2$ measured for the other fixation density. As the following equation shows, this is an estimate for the negative of the cross-entropy of the two densities $H(p_2;p_1)$.

\begin{align}
H(p_2;p_1) &= -\int p_2(x) \log(p_1(x))dx \\
&= -E_{p_2}\left(\log(p_1(x))\right)\\
&\approx -\frac{1}{n}\sum_{i=1}^n  \log\left(\hat{p}_1(f_2^{(i)})\right)
\end{align}

This cross-entropy is closely related to the Kullback–Leibler divergence $KL(p_2||p_1)$, which is simply the cross-entropy minus the entropy of $p_2$, i.e.,
\begin{equation}
KL(p_2||p_1) = H(p_2;p_1) - H(p_2)
\end{equation}
 
Thus, the log-likelihood we report measures how well $p_1$ approximates $p_2$, irrespective of the entropy of $p_2$, i.e., irrespective of the upper limit for predictions of $p_2$. 

To implement this, we again used leave-one-subject-out cross-validation, i.e. for each subject we computed a separate kernel density estimate $\hat{p}_1$ using only data of the other subjects and evaluated it at the fixation locations of that one subject.


\paragraph{Comparisons over Time.} 
Specifically, we compare fixation densities over time taking the distributions of fixations with two given fixation \#s as $p_1$ and $p_2$,i.e., we measure how (dis-)similar fixation densities with different fixation \#s are. For the kernel density estimates necessary in this computation we tested bandwidths from 1.0 to 5.0 dva in steps of 0.2 dva. We report the results with a bandwidth of 1.6 dva, which results in the highest likelihood for fixation \#s 2-25 averaged over all predicting fixation \#s, images and subjects.

\paragraph{Comparisons between Targets.}
For the search data we compared the fixation densities produced by subjects when searching for different targets on the same images. For this comparison we tested bandwidths from 0.5 to 1.5 dva in steps of 0.05 dva and report values computed with a bandwidth of 0.9 dva, which results in the highest likelihoods averaged over all images, comparisons and subjects

\subsection{Evaluation of saliency models}

In our analysis of saliency models we largely follow \cite{kummerer2015}, who recommend to use the log-likelihood of fixations under the model for evaluation after fitting a non-linearity, blur and center bias for each model to map the saliency map to an optimal prediction for the fixation density. Some transformations cannot be avoided, because classical saliency models do not predict a fixation density, but only a saliency map, which is not necessarily a density. Also, saliency maps only aim to be monotonically related to the fixation density when averaged over patches. Thus, fitting a local non-linearity and a blur allows a fairer comparison between models by fitting the parts of the model, which matter for the evaluation based on likelihoods, but not necessarily for the criteria used to design the models. Furthermore, we were interested in how well the fixation density can be predicted with certain predictors, which also argues for an optimal mapping from saliency map to fixation density.

\begin{figure*}
\includegraphics[width = \textwidth]{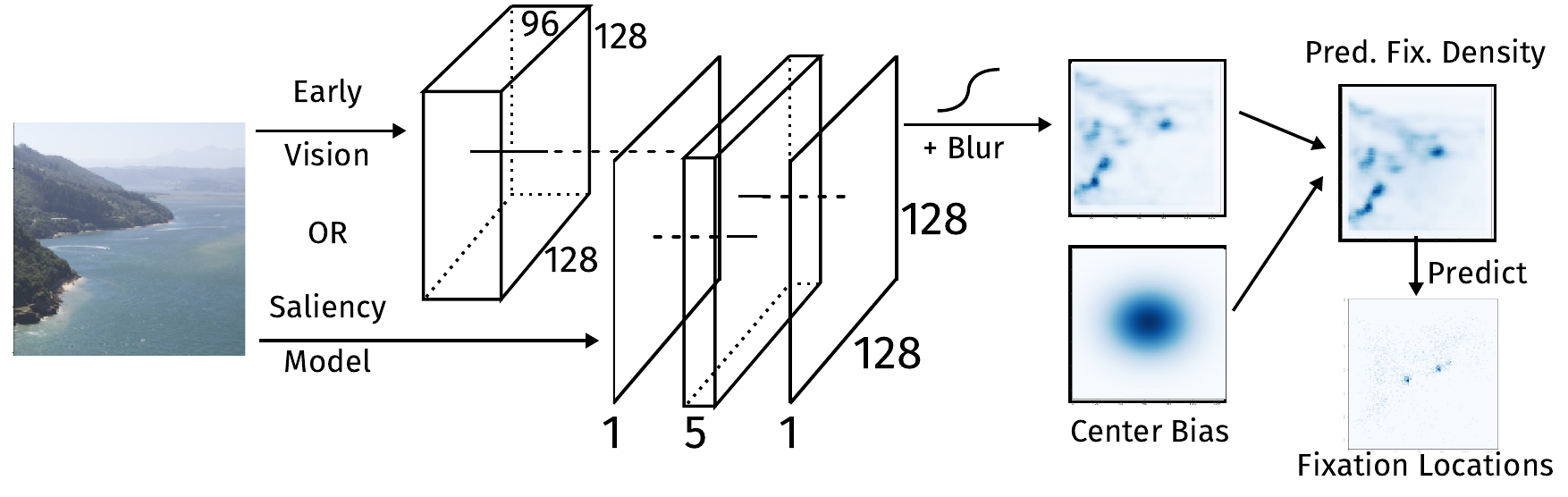}
\caption{Shallow neural network to map raw saliency models to fixation densities. We first compute a raw saliency map from the image, either by applying the saliency model or by linearly weighing the 96 response maps produced by our early vision model. Then two $1\times1$ convolutions are applied which first map the values to 5 intermediate values per pixel locally and then map to a single layer with a Relu non-linearity inbetween, which effectively allows a piecewise linear map with 5 steps as an adjustable local non-linearity. We then apply a fixed sigmoidal non-linearity and blur with a Gaussian with adjustable size. Finally we multiply with a fitted Gaussian Center Bias, which results in the predicted fixation density, which can be evaluated based on the measured fixation locations.\label{fig:DNN}}
\end{figure*}

To fit the mapping from the saliency map to the fixation density, we used the deep neural network framework Keras as included in TensorFlow \citep[version 1.3.0]{tensorflow2015} as a backend. In this framework, we fit a shallow network as illustrated in Figure \ref{fig:DNN} for each saliency model separately after resizing the saliency maps to $128\times128$ pixel resolution and rescaling the saliency values to the interval $[0,1]$. 

The network contained two conventional $1\times1$ convolution layers which first map the original to an intermediate layer with 5 channels and then to a single output layer, allowing for a broad range of strictly local non-linear mappings to the fixation density. 

Next, we apply a blurring filter to the activations, which allows saliency to attract fixations to nearby locations. This is not to be confused with blurring the original image, which has entirely different effects on saliency computations, changing the features available to the saliency models for example. To implement the blur we used a $25\times25$ custom convolution layer, in which we set the weights to a Gaussian shape of which we fitted the two standard deviations.

Finally, we apply a sigmoidal non-linearity to map activations to a strictly positive map and apply a center bias through a custom layer. This layer first multiplies the map with a Gaussian with separately fitted vertical and horizontal standard deviations and then normalises the sum of the activities over the image to 1 to obtain a probability density.\footnote{We also implemented an additive center bias, which performed worse than the multiplicative version for all models.}

As a loss-function for training the network, we directly use the log-likelihood as for the kernel density estimates described above. In Keras we implemented this by flattening the final density estimate and using the standard loss function \emph{categorical\_crossentropy} to compare to a map with sum 1 and entries proportional to the number of fixations at each location.

For evaluation, we performed fivefold cross-validation over the used images, i.e., we trained the network 5 independent times leaving out one fifth of the data. For training we ran the the Adam optimisation algorithm \citep{kingma2014} with standard parameters till convergence by reducing the learning rate by a factor of 2 whenever the loss improved less than $10^{-5}$ over 100 epochs and stopping the optimisation when the loss improved by less than $10^{-6}$ over 500 epochs. We did not employ a test set here as we did not optimise any hyper parameters and did not use any stopping or optimisation rules based on the validation set.

\subsection{Interpretation of the log-Likelihood Scale}
We measure all model performances in $\frac{bit}{fix}$ relative to a model that predicts a uniform distribution of fixation locations. A $\Delta$ log-likelihood of $0 \frac{bit}{fix}$ is equal to a uniform prediction over the image---thus no gain over predicting that any position in the image is equally likely to attract a fixation. 
If a model is $1 \frac{bit}{fix}$ better than a uniform model, its density is on average $2$ times higher than the density of the uniform model. As the density has to integrate to one this corresponds roughly to reducing the possible area for fixations by the same factor, i.e., in the case of $1 \frac{bit}{fix}$ to half the size. Thus, larger $\frac{bit}{fix}$ result from an increased accuracy of the prediction. In general this relation is $2^x$ for a difference of $x\frac{bit}{fix}$. The interpretation of a difference in log-likelihood per fixation as a factor on the density or predicted area is independent of the absolute performance of the compared models, i.e., if one model already reaches a performance of $2 \frac{bit}{fix}$ another model reaching $3\frac{bit}{fix}$ still predicts densities which are twice as high at fixation locations and thus restrict each fixation to roughly half the area.

\subsection{Tested saliency models}

To get a comprehensive overview over saliency model performance, we chose a few representative models for predicting saliency:

\paragraph{Kienzle.} As an example of an extremely simple low-level model of visual saliency, we employ the model by \cite{kienzle2009}, using the original implementation supplied by Felix Wichmann. 

\paragraph{Itti \& Koch.} As the most classic saliency model, we evaluate the original model by \cite{itti1998}. To compute the saliency maps we used the implementation which accompanies the GBVS saliency model, which performed decisively better than the original implementation from www.saliencytoolbox.net.

\paragraph{GBVS.} As a better performing classical hand crafted saliency model, we use the Graph based visual saliency model by \cite{harel2006}. Code was downloaded from \textcolor{blue}{\href{http://www.vision.caltech.edu/{\textasciitilde{}}harel/share/gbvs.php}{here}}\footnote{Link for the paper version: http://www.vision.caltech.edu/\textasciitilde{}harel/share/gbvs.php}.

\paragraph{DeepGaze II.} As a representative of the newest deep neural network based saliency models, we chose DeepGaze II by \cite{kummerer2016}. This model is currently leading the MIT-saliency benchmark \citep{mit-saliency-benchmark}. Saliency maps for this model were obtained from the webservice at deepgaze.bethgelab.org as log-values in a .mat file and converted to linear scale before use.

\paragraph{Early Vision.} Our early vision saliency model is based on our psychophysical spatial vision model we published recently \citep{schutt2017b}. This model implements the standard model of early visual processing to make predictions for arbitrary luminance images. As an output it produces a set of $8 \times 12 = 96$ orientation $\times$ spatial frequency channel responses, spatially resolved over the image. 

To obtain a saliency map from these channel responses we linearly weigh and added them to form a saliency map. To map this sum to a predicted fixation density, we used the same machinery as the saliency maps for all other models. 

The weights for the initial sum were unknown, however, and needed to be fit. Fortunately, we could implement the weighted sum as a $1\times1$ convolution layer in TensorFlow as well. This implementation allows us to interpret the weights as additional parameters of the mapping to the fixation density. Thus we could train an arbitrary weighting for the maps from our early vision model directly while keeping the benefits of a non-linearity, blur and center bias as for the other models. 

\section{Results: Scene Viewing}

\subsection{Overall saliency model performance}

\begin{figure*}
\includegraphics[width = \textwidth]{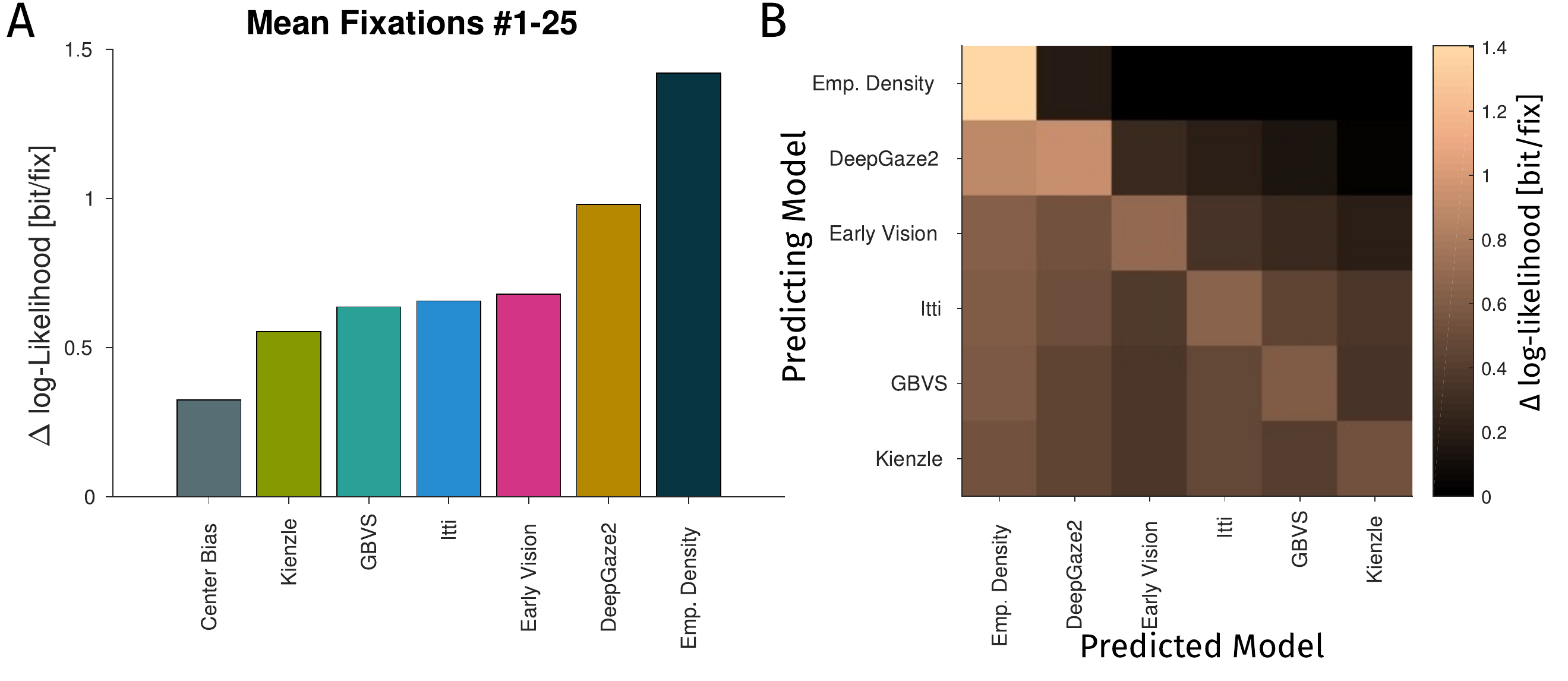}
\caption{A: Average performance of the models. B: Similarity of the different saliency maps. Measured in terms of $\Delta$ log-likelihood, i.e., as the prediction quality when using one map to predict random draws from another.  \label{fig:CorpusSaliency}}
\end{figure*}

Before we analyse the temporal evolution of the fixation density, we compare the overall performance of our saliency model based on a model of spatial vision to a range of classical low-level saliency models, i.e., Itti \& Koch \citep{itti1998}, GBVS \citep{harel2006}, Kienzle \citep{kienzle2009}, and the currently best DNN-based saliency model DeepGaze II \citep{kummerer2016}. To make the models comparable, we fitted the same non-linear map, blur and center bias for all models (see Methods). As the evaluation criterion, we use the average log-likelihood difference to a uniform model as described by \cite{kummerer2015} for saliency models and \cite{schutt2017} for dynamical models (see Methods).

The results of the overall analysis\footnote{We restrict ourselves to fixations \#1-\#25 here for consistency with later plots over time. Including later fixations does not qualitatively change the results displayed here. The only change is that later fixations are predicted worse by all models decreasing the absolute performance of all models slightly.} are displayed in Figure \ref{fig:CorpusSaliency}. All low-level saliency models predict fixations better than a pure center bias model. Our early vision based saliency model performs slightly better than the classical saliency models using only a simple weighted sum of activities as a saliency map. Thus, a simple sum of activities of a realistic early spatial vision model seems to be sufficient for modelling low-level influences. 

DeepGaze II clearly outperforms all tested classical saliency models \citep{itti1998, kienzle2009, harel2006} and our early vision based model by $0.3 \frac{bit}{fix}$. But DeepGaze II is not as close to perfect prediction for our scene viewing dataset as for the MIT-saliency-benchmark, missing it by roughly  $0.4 \frac{bit}{fix}$ \citep[compare our Fig.~\ref{fig:CorpusSaliency}A to][Fig.~3]{kummerer2016}. A potential reason for this might be that our dataset contains many more fixations per image ($\approx 2600$), than the saliency benchmark \citep[39 observers $\times$ 3 seconds $\leq 390$]{judd2012}, which allows a more detailed estimation of the empirical fixation density. An alternative but not exclusive explanation is that the MIT-saliency-benchmark dataset contains (more) humans, faces and text, which might help DeepGaze II, as these are typical high-level properties reported to attract fixations.

These overall performance results suggest that a realistic early vision representation provides similar predictive value for the density of fixations as classical saliency models do. The results do not fully answer the question whether classical saliency truly represents early visual processing though. To approach this question, we additionally analysed the similarity of predictions of all saliency models. To compare saliency model predictions, we calculated their performance in predicting each other on the same log-likelihood scale we use to compare how well they predict human fixations.

The resulting cross-entropies between saliency models are shown in Figure \ref{fig:CorpusSaliency}B. Each cells' colour indicates how well the density created by one model (predicting model) predicts draws from another model's density (predicted model). We first look at the diagonal, which represents how well each model predicts itself, i.e., the entropies of the different model predictions. The empirical density predicts itself more accurately than any saliency model predicts itself. Also, each of the saliency models is distinct from the others, as the diagonal elements have larger values than any corresponding off-diagonal ones.

Next, we can observe some asymmetry in the prediction qualities, as the models are sorted according to their prediction quality. Generally, the top right quadrant is darker than the lower left quadrant, i.e., well-performing models predict poor performing models less accurately than poor performing models predict well-performing models. For example, the empirical fixation density is predicted reasonably well by all saliency models, but is itself not a good predictor of saliency maps. This pattern indicates that even poorly performing models predict density where people fixate. However, they seem to add density where at locations which are not fixated by humans and not predicted by more successful models such better models reject locations more efficiently.

The tendency that more successful saliency models generally become more specific than less successful ones is partially caused by the link to the fixation density we fit. The local non-linearity allows the model to adjust how strongly the prediction of the model is weighted, i.e., how large the difference between the peaks and valleys of the prediction is. To optimise their performance, weaker models can use this mechanism to increase their predicted density in the valleys, as a substantial proportion of human fixations fall into these valleys. This mechanism broadens the prediction of weak models more than that of strong models. 

Finally, there is a group of models which make less dissimilar predictions: Low-level saliency models share some common entropy, i.e., the off-diagonal values for these are higher than between other models. Especially, classical models predict each other better than our new early vision based model. These results imply that the early vision based saliency is somewhat different from classical saliency models.

\subsection{Predictability of fixation densities over time}
\begin{figure*}
\includegraphics[width = \textwidth]{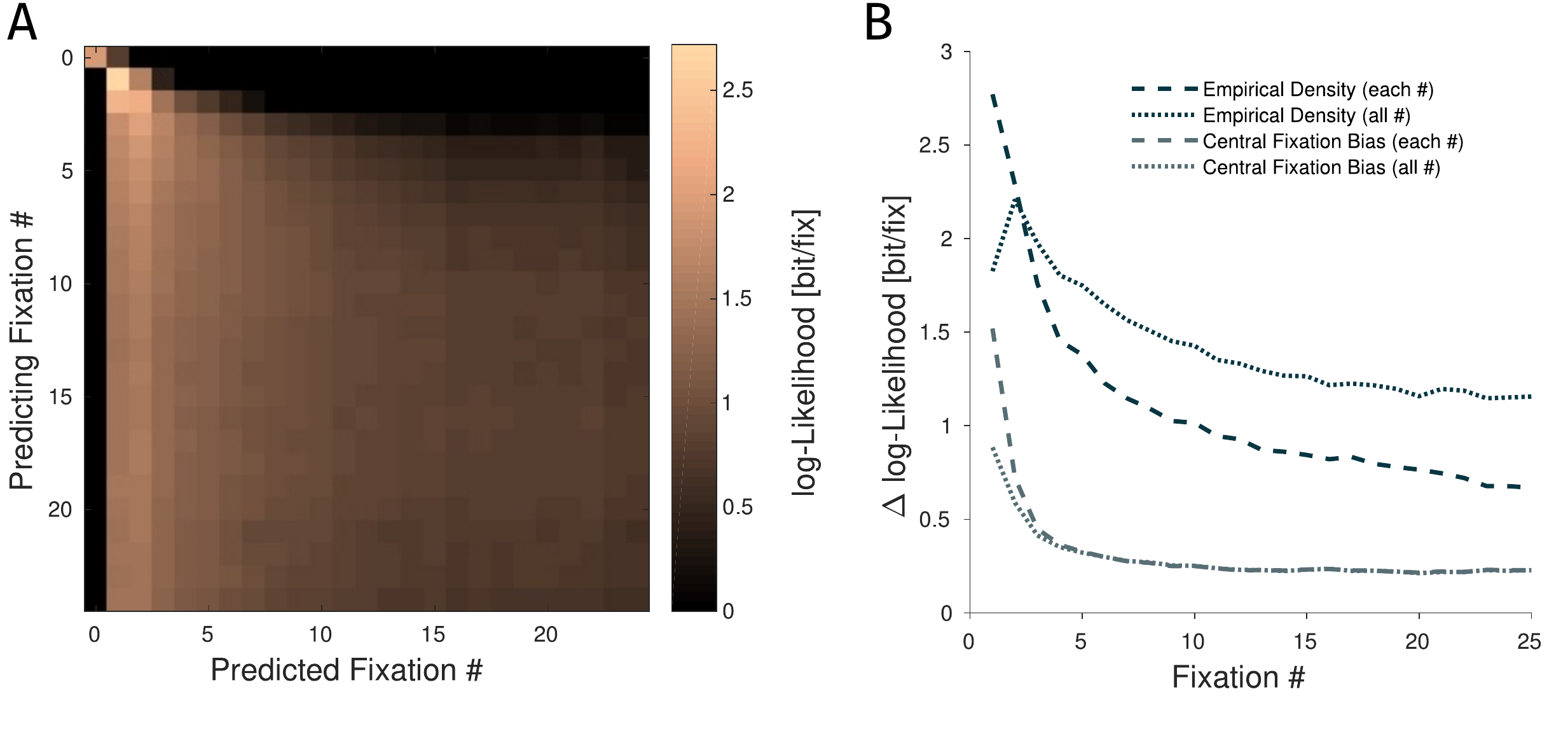}
\caption{Analysis of the predictability of fixation densities over time. A: log-likelihood for predicting the fixations with one fixation number from fixations with a different fixation number as a measure how well the density at one fixation number predicts the fixations with a second fixation number. B: Performance of the Gold Standards over time. Shown are the performance of the empirical density measured by predicting the fixations of one subject from the fixations of other subjects and the central fixation bias measuredd by predicting the fixations in one image based on the fixations in other images. For each of these limits two curves are shown: One continuous line based on only fixations with this fixation number and one dashed line based on all fixation numbers. \label{fig:CorpusTime}}
\end{figure*}

To evaluate saliency models over time, we computed log-likelihoods for each fixation \# of each trial on an image. From these fixations, we computed a kernel density estimate and evaluated likelihoods for each fixation number using leave one subject out cross-validation. The results of this analysis averaged over images are displayed in Figure \ref{fig:CorpusTime}A. Different rows correspond to using different fixations for constructing the prediction. Different columns correspond to predicting different fixations with a trial. Caused by the cross-validation over subjects, the estimates for a fixation number predicting itself are interpretable and comparable to the predictions for other fixations.

Going through the plot in temporal order we find that: (i) The 0th fixation (the starting position) neither predicts the other fixation locations nor is predicted by them well, which was to be expected, since the starting position is induced by the experimental design. (ii) The first and, to a lesser degree, the following fixations show an asymmetric pattern: They predict other fixations badly, but are predicted well by higher fixation numbers, indicating that they land at positions which are fixated later as well, but do not cover all of them. (iii) This tendency gradually declines from the second fixation till roughly fixation \#10, accompanied by a gradual decline in predictability. (iv) From fixation \#10 onwards the fixation densities of all fixation numbers predict each other equally well, indicating that the fixation density has reached an equilibrium state. 

These results suggest a separation into three phases: (i) The first fixation, which seems to be different from all others, (ii) the phase with the asymmetric pattern when fixations are well predicted by the later density but have not converged to it yet, and (iii) the final equilibrium phase when the fixation density has converged. 

Our next aim was to quantify the maximum amount of image-based predictability at different time points after image onset. To quantify upper and lower bounds, we used four limiting cases: First, a central fixation bias, implemented as a kernel density estimate from fixations with the same fixation number from all trials on all images. Second, a central fixation bias based on all fixations from all images. Third, the empirical density estimated as a kernel density estimate from the fixations with the same fixation number on the same image. Fourth, a different estimate of the empirical density estimated from fixation \#2 to fixation \# 25 on the given image, to increase the number of fixations available for the kernel density estimation. All four estimates were again calculated using leave one subject out cross-validation, such that only fixations from other subjects were used for estimating the density.

The results of this analysis are displayed in Figure \ref{fig:CorpusTime}B. The central fixation bias declines quickly from the good prediction based on the initial central fixation bias on the first fixation to a constant level of roughly $0.25 \frac{bit}{fix}$, which is retained over the remaining trial. Also, the two estimates of the central fixation bias only differ substantially for the first few fixations, affected by the initial central fixation bias.

For the empirical density, both estimates show a gradual decline over time. The estimate based on all fixation numbers flattens out between fixation \# 10 and fixation \#15. The estimate based only on fixations with the same fixation number quickly falls below the estimate based on all fixations and keeps decreasing. This trend is most likely due to the lower and decreasing number of fixations. As trials always lasted 10 seconds and fixation durations vary, the number of fixations within a trial varies substantially. Thus, we have fewer examples for high fixation numbers ($\sim$~66\% for fixation~\#25). First fixations are better predicted using only other first fixations despite their smaller count, confirming that the first fixation follows a different density than later ones.

We interpret this observation as further evidence for a separation into a short initial period with a strong initial central fixation bias, a period for which predictability gradually declines and a late equilibrium period. Additionally, the difference between our two estimates of the maximally predictable information shows that the $\sim100$ fixations we have for each fixation number are not enough for a good estimate of the fixation density. Thus the fixation density estimate from all later fixations on an image gives a better estimate of the maximally attainable fixation density for all but the first and possibly second fixation. The initial fixations (fixation \#1 -- \#2) seem to deviate from what attracts later fixations. 

\subsection{Influence of low- and high level features over time}

\begin{figure*}
\includegraphics[width = \textwidth]{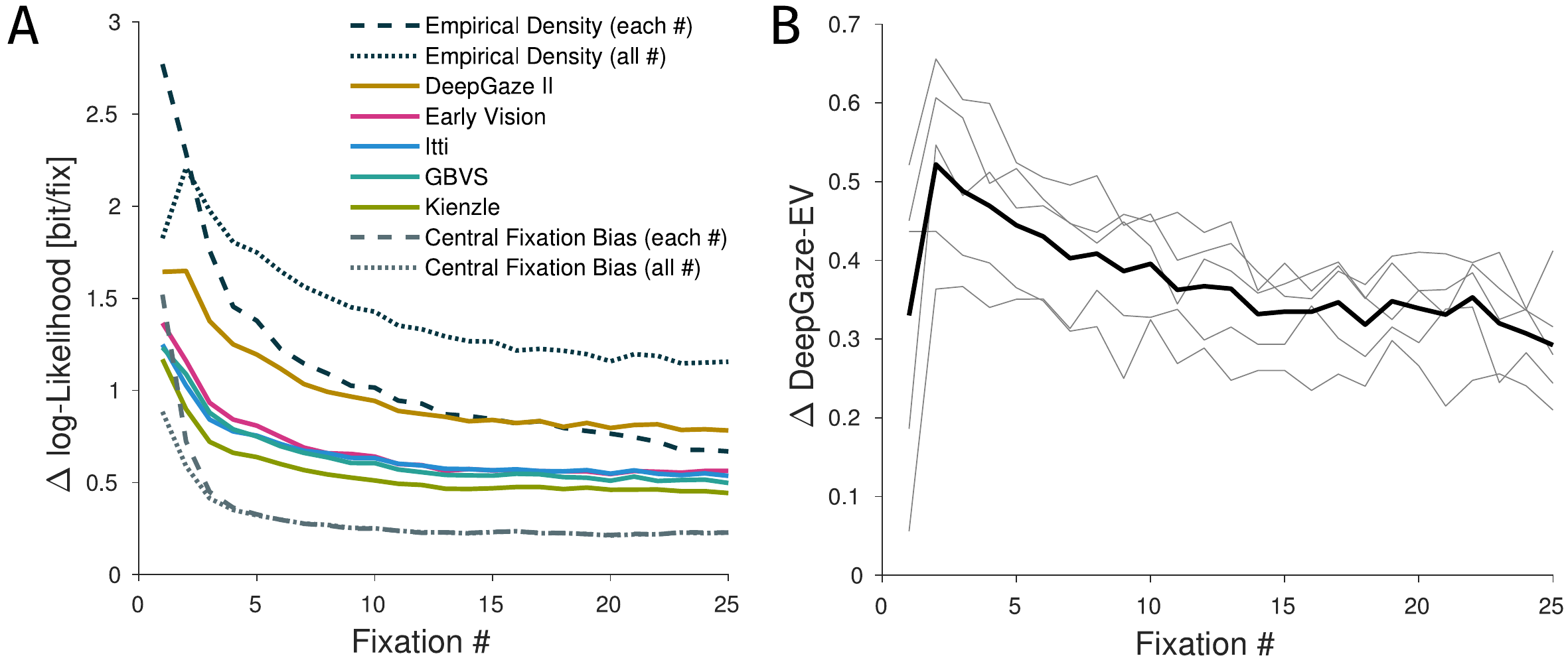}
\caption{Saliency model performance on the scene viewing dataset. A: Performance of the saliency models over time, replotting the maximal achievable values from Figure \ref{fig:CorpusTime}. B: Difference between DeepGaze II and the early vision model over time. The grey lines represent the individual folds.   \label{fig:CorpusTimeSaliency}}
\end{figure*}

We are interested in the performance of saliency models over time to test whether low-level features play a more important role at the beginning of a trial. The results of this evaluation are displayed in Figure \ref{fig:CorpusTimeSaliency}A. In general, prediction quality of all saliency models follows the curve for the empirical density with a gradual decline that reaches a plateau between fixation \#10 and \#15. As expected, all saliency models are better than the central fixation bias, as our fitted mapping includes a central fixation bias, but do not perfectly predict the empirically observed fixation densities. 

Differences between models in their overall performance are present throughout the trial. DeepGaze II performs best and other saliency models run largely in parallel about $0.3-0.5\frac{bit}{fix}$ below.
To investigate the additional contribution of high-level features, we plot the difference between DeepGaze II and the early vision based model in Figure \ref{fig:CorpusTimeSaliency}B. This plot emphasises that DeepGaze II is constantly predicting fixations better than the early vision based model (all lines are always $\gg0$). This difference is especially large during the initial exploration phase during which the advantage follows the general decline in predictability (fixation \#2 -- \#10). In contrast, the advantage of DeepGaze II for the first fixation (\# 1) is as small as for fixations during the equilibrium phase ($\approx 0.3\frac{bit}{fix}$), which results in a substantial jump from fixation \#1 to fixation \#2, where DeepGaze II has the largest advantage of $\approx0.5\frac{bit}{fix}$. As the first fixation contains a strong central fixation bias, which varies over the time course of trial \citep{rothkegel2017}, and was proposed as the main point in time for low-level, bottom-up effects \citep{anderson2015}, we analyse this first fixation in more detail. 

\subsection{Density of the first fixation}

\begin{figure*}
\includegraphics[width = .5\textwidth]{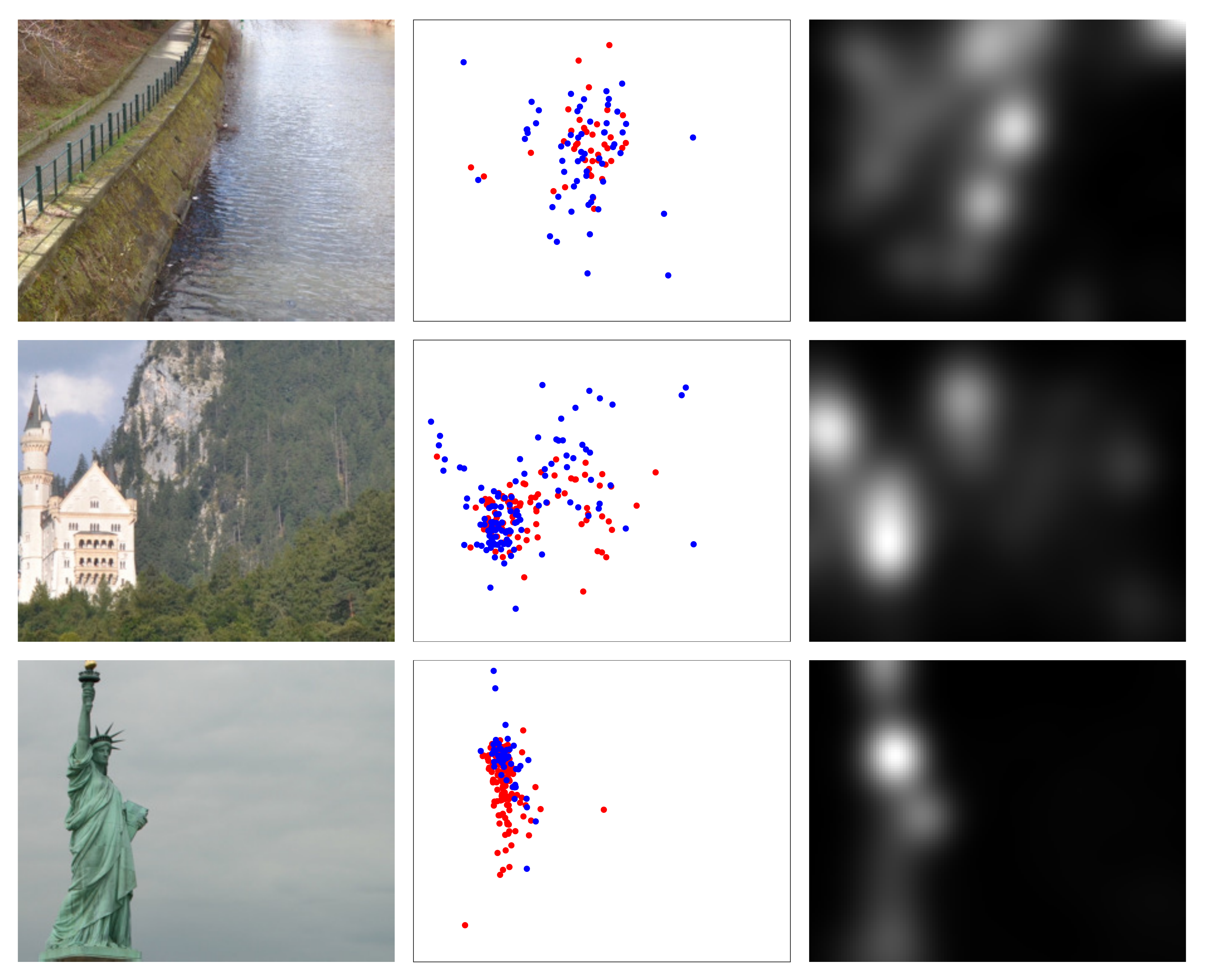}
\includegraphics[width = .5\textwidth]{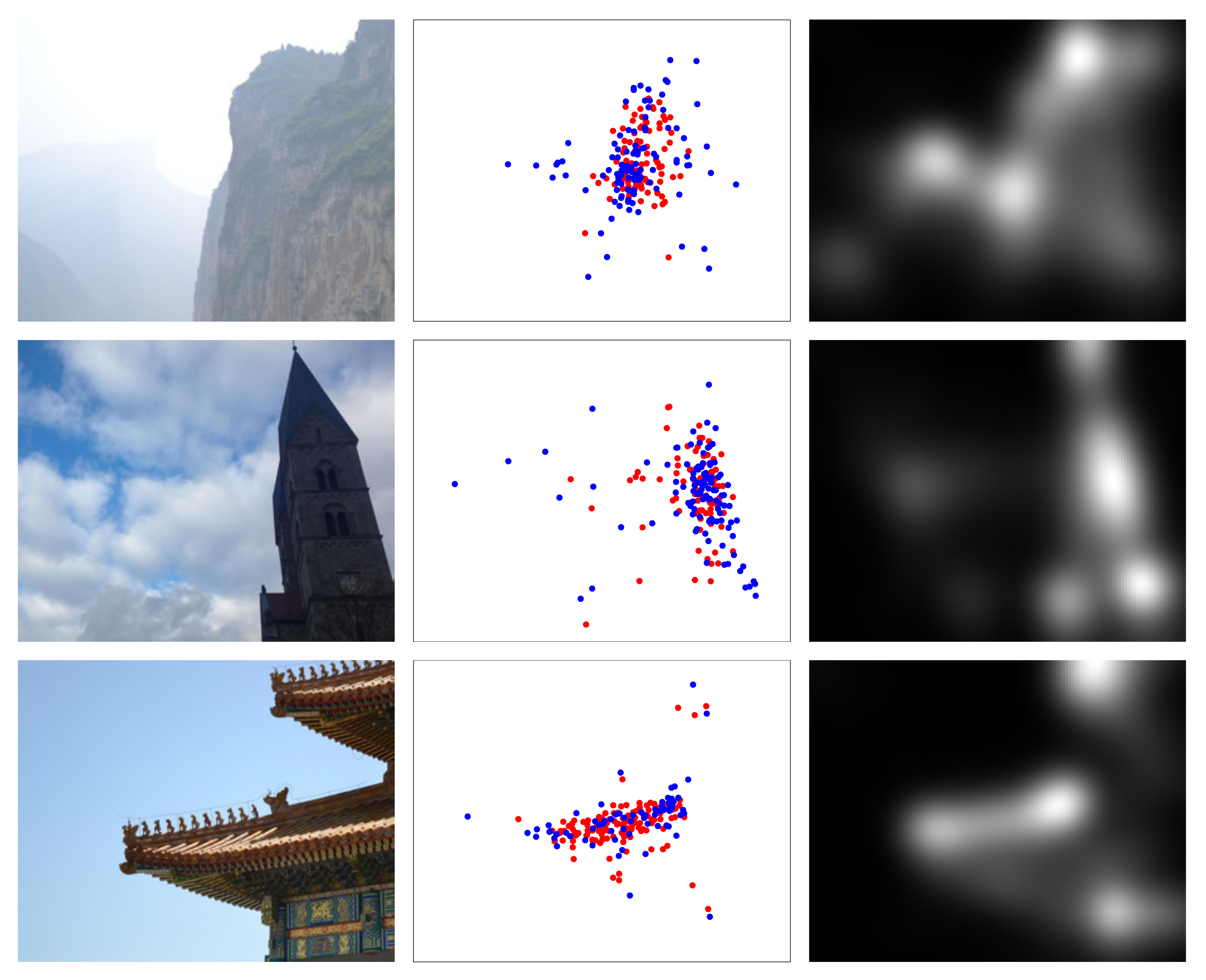}
\caption{Examples showing the differences among images in the initial central fixation bias. For each image we show the image, the first chosen fixations as a scatter plot and the density of all later fixations. Colour represents a median split by the fixation duration at the start location, red fixations were chosen after less than $270ms$, blue fixations after more than $270ms$. The left column shows examples of our left focussed images, the right column of the right focussed ones.  \label{fig:CenterbiasDetail}}
\end{figure*}

\begin{figure*}
\begin{center}
\includegraphics[width = \textwidth]{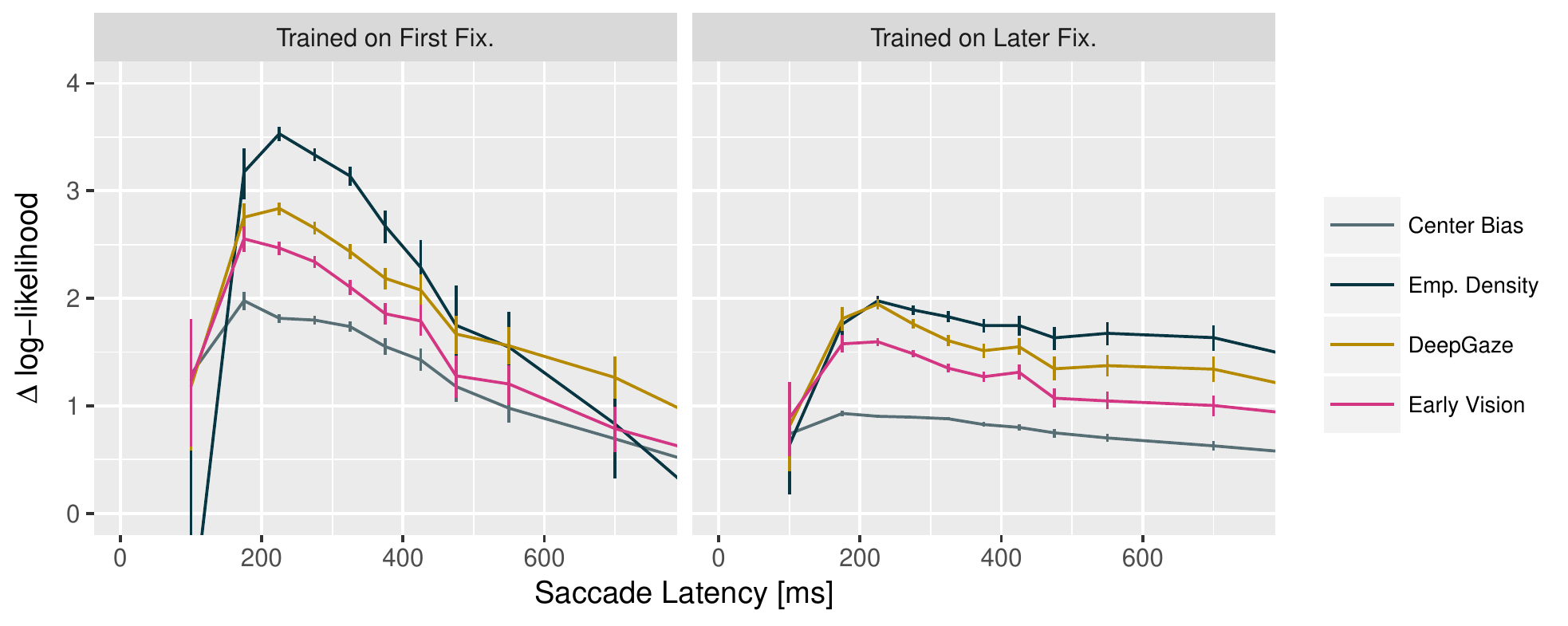}
\end{center}
\caption{Temporal evolution of prediction qualities for the first fixations against the latency of the saccade towards them. We plot the log-likelihood gain compared to a uniform distribution for empirical density, center bias, early vision based saliency model and DeepGaze II. For display saccade latencies were binned, errorbars represent bootstrapped 95\% confidence intervals for the mean. \label{fig:FirstFix}}
\end{figure*}

To analyse the first fixation in detail, we performed two complementary analyses: First, we display the first fixation location of participants on some example images in Figure \ref{fig:CenterbiasDetail}. Second, we split the data from the first fixation by the latency of the first saccade after image onset. This split allows us to compare the performance of our early vision based model and DeepGaze II to the performance of the center bias and the empirical density prediction depending on onset of the first saccade. For each predictor we created two separate fits: One based on only first fixations and one based on all but the first fixation (fixation \#2 -- \#25). For the saliency models we retrained our network, i.e., learned a separate blur, non-linearity and center bias. For the empirical density and center bias we generated separate kernel density estimates. The results of this second analysis are plotted in Figure \ref{fig:FirstFix}.

Generally the density of the first fixation shows a pronounced \emph{initial} center bias \citep[c.f. Fig.~\ref{fig:CorpusTime}, ][]{rothkegel2017, tatler2007}, i.e., early saccades almost exclusively move towards the center of an image. This tendency is visible in the raw data (for example in the upper left image in Fig.~\ref{fig:CenterbiasDetail}) and in the high prediction quality of the image independent central fixation bias model for the first fixation (see Fig.~\ref{fig:FirstFix}, light grey line). A potential cause of the central fixation could either be that a certain proportion of fixations is  placed near the image center independent of image content or that fixation locations depend on image content weighted by the distance to the center. However, exploring first fixations in more detail shows at least two problems with these simple accounts, illustrated by the examples in Figure \ref{fig:CenterbiasDetail}. First, the strength of the central fixation bias differs considerably between images. For some images fixations are indeed consistent with a Gaussian distribution around the image center (e.g., top left). For other images fixations locations seem to stem from a mixture of a Gaussian distribution and a distribution depending on image content (e.g., left middle) or are strongly dominated by image content with almost no fixations near the center (e.g., bottom left). Second, when first fixations depend on image content, the distribution of first fixations differs from the distribution of later fixations in some images (e.g., right bottom), where the distribution of first fixations shows a different peak than later fixations. Thus the interaction of central fixation bias and image content seems to be more complex than a simple additive or multiplicative relation.

In addition to the central fixation bias, we observe that first fixations are clearly guided by image content. We find that fixations can be predicted much better, when knowledge about the image is included (see Fig.~\ref{fig:CorpusTime} \& Fig.~\ref{fig:FirstFix}, difference between empirical density and central fixation bias) and can confirm this by looking at examples in Figure \ref{fig:CenterbiasDetail} (distributions clearly differ between images and depend on identifiable objects in images). We can also confirm the observation that the first fixation differs from later fixations as all predictions fitted to the first fixation perform much better than predictions fitted to later fixations (compare left \& right plot in Fig.~\ref{fig:FirstFix}). This benefit is visible in the raw data as the distribution of first fixations generally deviates from the density computed from later fixations (see Fig.~\ref{fig:CenterbiasDetail}, compare scatter plot to fixation density). 

The image guidance is captured by saliency models to some extent. Low-level features as computed in the early vision based model perform $0.4\frac{bit}{fix}$ and $0.5\frac{bit}{fix}$ better than the central fixation bias for training based on the first and later fixations respectively, i.e., its predicted density is 1.34 and 1.44 times higher at fixation locations than the density of the central fixation bias. In addition, DeepGaze II performs $0.3\frac{bit}{fix}$ better than the early vision based model, i.e., its predicted density is on average $\approx 1.25$  times higher than the early vision based model. Thus, high-level features predict fixation locations better than low-level features already for the first fixation. These differences are comparable to later fixations, but all estimates are much higher than for later fixations due to the central fixation bias already explaining $1.6\frac{bit}{fix}$ and $0.85\frac{bit}{fix}$ respectively, i.e., its density is already up to 3 times as high at first fixation locations than the uniform distribution.

Analysing the effect of onset time of the first saccade (saccade latency), all predictions are relatively bad for latencies below $150 ms$. These fixation locations appear not to be guided by the image, but represent only 5\% of first fixations. After this poor performance follows the bulk of fixations between 200 and 400 ms which are best predicted by all models. For these fixations the early vision model performs up to $0.7\frac{bit}{fix}$ better than the central fixation bias, but the DeepGaze II model is consistently $\approx0.3\frac{bit}{fix}$ better than the early vision model. After this, we see a decline in prediction quality of the models trained for the first fixation emphasising that late saccades follow a different density than earlier ones. The models trained on the later fixations decline much slower. This slower decline of models trained on later fixations could be the earliest part of the general decline in predictability we observe over multiple fixations above. Thus, fixations after a long first saccade latency might already follow the same factors as subsequent fixations.

Interpreting these results, we conclude that high-level information is advantageous for the prediction of eye movements already 200ms after image onset. However, the central fixation bias and low-level guidance are much better models for the first fixation than for later ones, especially for relatively early saccades. 

\section{Results: Visual Search}

\subsection{Predictability of fixation densities for different targets}
\begin{figure*}
\includegraphics[width = \textwidth]{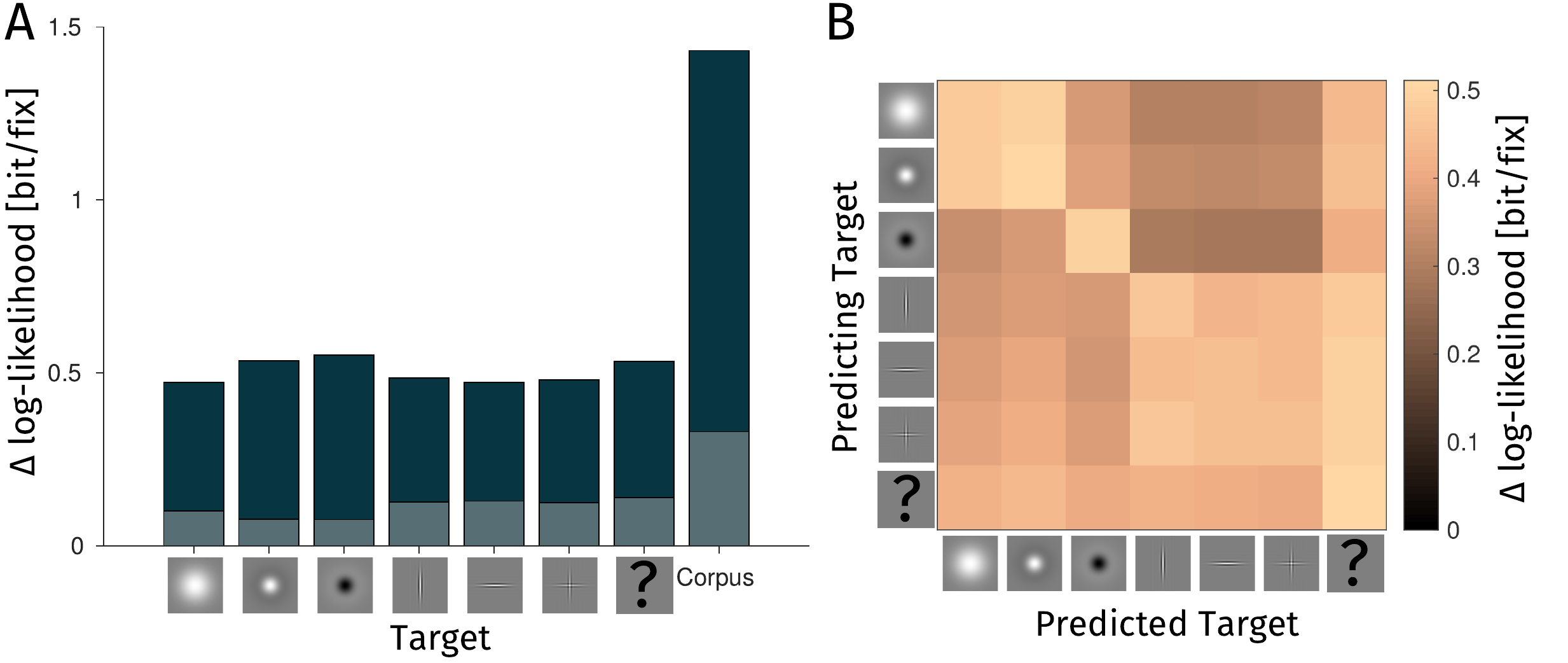}
\caption{Analysis of fixation densities in the search experiment. A: Prediction limits for the fixation densities for the different search targets estimated from leave one subject out cross-validation. The grey lower proportion indicates the maximum for image independent prediction (central fixation bias). The black bars represent the maximum for image (\& target) dependent prediction. We additionally plot these values for the scene viewing dataset for comparison. B: $\Delta$ log-likelihood as a measure of prediction quality when predicting the fixation locations when searching for one target from the fixation locations when searching for a different target in the same image. \label{fig:SearchGold}}
\end{figure*}

The first analyses of the visual search data examine whether fixation locations are predictable from the image and if fixation densities differ for different search targets. To investigate this, we calculated kernel density estimates from the fixation locations for each search target. We evaluate how well these kernel density estimates predicted the fixations made while subjects searched for the same or other targets using the same $\frac{bit}{fix}$ likelihood scale we use for model evaluation, which estimates the (cross-) entropies of the fixation distributions in this case (see Methods).

The results are displayed in Figure \ref{fig:SearchGold}. In panel A we plot the performance of the empirical density (black bar) and the central fixation bias (grey bar) of fixation densities for different targets. These estimates were computed the same way as for the scene viewing dataset and are based on a similar number of fixations per image. Comparing these likelihoods to the scene viewing data reveals that fixation locations during visual search are distributed much broader over the images than in the standard scene viewing task. Depending on the target the fixation density contains only $0.5-0.6\frac{bit}{fix}$ of predictable information. In contrast, the empirical density in the scene viewing data explained $\approx 1.4\frac{bit}{fix}$. 

In panel B we display how well the fixation distributions for the different targets predict each other. The fixation distributions all predict each other to some extend ($>0.3\frac{bit}{fix}$ for all pairs). Furthermore, the fixation densities of our targets separate into three groups of targets whose fixation distributions predict each other roughly as well as themselves indicating practically identical fixation distributions. The three high spatial frequency targets lead to similar fixation distributions and the Gaussian blob and the positive Mexican hat lead to similar distributions, while the negative Mexican hat produces a different distribution from all others. That the distribution is so different for the two polarities of the Mexican hat is somewhat surprising from the perspective of early spatial vision, as these stimuli have equal spatial frequency content. Thus, this finding might hint at a greater importance of differences between on and off channels in pre-cortical processing \citep{whittle1986}.

Nonetheless, log-likelihoods for the fixations of any target were higher under the fixation densities estimated for any other target than for the uniform distribution (all cells $\gg 0$). This indicates that some areas attract fixations independent of the search target. 

In summary, our results show that fixation locations can be predicted to some extent, although fixations are distributed much broader than in the scene viewing experiment\footnote{As we did not use the same images for the two experiments we cannot rule out an effect of image content entirely. However, 87 of the 90 images in the scene viewing experiment had higher average likelihoods than the best predictable image from the search experiment, i.e., the distributions were almost non-overlapping. Thus this explanation is unlikely to explain the whole effect.}. While there is some overlap across fixation locations for different targets, fixation locations also depend on specific target features. This corroborates our earlier observation that searchers adjust their eye movements to the target they look for \citep{rothkegel2018}.

\subsection{Influence of low- and high-level features over time}

For the analysis of the saliency models to investigate the role of feature complexity,  we employed the same techniques as for the scene viewing dataset. We fit a non-linearity, blur and central fixation bias and evaluate the performance of the resulting prediction over time using cross-validation.

\begin{figure}
\includegraphics[width = \textwidth]{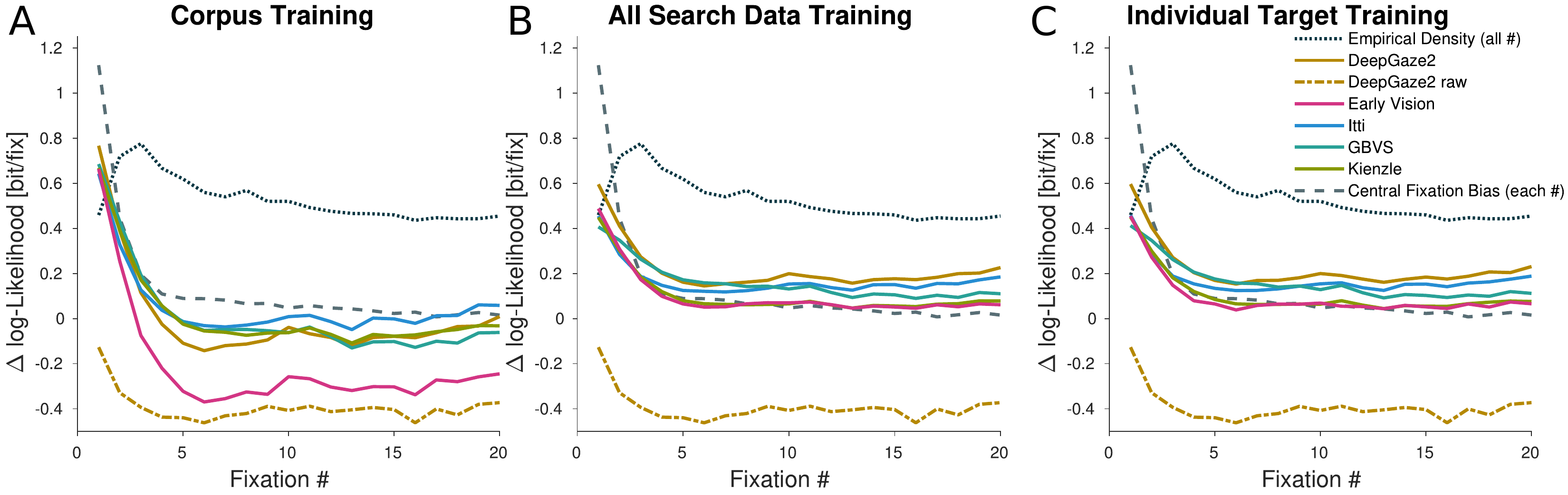}
\caption{Performance of the saliency models on the search dataset over time. The different columns show different conditions for training the connection from saliency map to fixation density. Free-viewing training: taking the mapping we trained for the scene viewing experiment. All search data training: Using all search data from the training folds. Inidividual target training: Training and evaluation was performed separately for each search target; We report the average over targets. Additional to the different saliency maps, we plot the empirical densities' performance (average over densities fit per target to fixations $\geq 2$), the center bias performance fitted for each fixation number and the performance of the unmodified DeepGaze II saliency map (DeepGaze2 raw). \label{fig:SearchSaliency}}
\end{figure}

As we show in Figure \ref{fig:SearchSaliency}, no saliency model predicts the fixation density well during visual search beyond the first few fixations. When we do not adjust the density prediction to the search data, the models are worse than a uniform prediction at most timepoints. The only time these densities predict fixation locations above chance are the first and second fixations with the initial center bias. When we train the connection from saliency map to fixation density newly for the search data, the saliency models still explain only a tiny fraction of the fixation density. Even DeepGaze II and the version of \citep{itti2001} provided with GBVS, which perform best, explain less than 0.2 $\frac{bit}{fix}$, i.e., they predict less than a third of the explainable information and increase the average density at fixation locations by a factor of 1.14 at best. Adjusting the link even stronger, we also trained the connection from saliency to fixation density separately for each target. Such an adjustment had little effect for any of the saliency models and the early vision based model did not profit from this adjustment either, although its performance changed slightly and at least improved on the training dataset (not shown).

Finally, we evaluated the DeepGaze II model---which performed best for free viewing---without the link we provided (shown as 'DeepGaze2 raw'). This evaluation is important as a test that our fitting scheme for the mapping to a density works properly for the search dataset as well. All other models do not predict a density map themselves. Thus, this evaluation is only possible for DeepGaze II, which already predicts a density as its saliency map. The raw prediction of DeepGaze II is clearly below chance performance, emphasising that the link we fitted here is not responsible for the failure of this model. 

Our results confirm that fixation locations during visual search are not predicted well by any bottom-up model \citep{najemnik2008,najemnik2009}. Neither high nor low-level features predict were humans look, whether they are adjusted to the task or not. 

\section{Discussion}
We explored the temporal dynamics of the fixation density while looking at natural images to investigate how low-level and high-level features and top-down and bottom-up control interact over the course of a trial. This analysis is made possible here for the first time by the long duration of trials and the large number of viewings for each image. 

\subsection{The temporal evolution of the fixation density}
Based on the similarities of fixation densities shown in Figure \ref{fig:CorpusTime} we suggest a separation of a typical scene-viewing trial into three phases:
\begin{enumerate}
\item An onset response which affects mostly the first saccade.
\item The main exploration, which is characterised by a gradual broadening of the fixation density.
\item A final equilibrium state, in which the fixation density has converged.
\end{enumerate} 

We interpret these three phases as an initial orienting response towards the image center, which can be biased by strong bottom-up signals in the image, followed by a brief guided exploration during which observers look at all parts of the image they are interested in and a final idle phase during which observers look around rather aimlessly.

Exploring the onset response in more detail, we found some guidance beyond a simple movement to the image center. An image dependent prediction performed substantially better than an image independent one (see Fig.~\ref{fig:FirstFix}). Examples of fixation densities for target of the first saccade (see Fig.~\ref{fig:CenterbiasDetail}) confirmed that fixations were guided by the scene sensibly with a bias towards the center.

The main exploration focusses on similar image locations as the subjects fixate, when the fixation density is converged (see Fig.~\ref{fig:CorpusTime}). The fixations during this phase are even better predicted by later fixation densities than later fixations themselves. During this phase the fixation density gradually broadens, becoming less and less predictable. Correspondingly, the performance of all saliency models is maximal at the beginning of this phase and decreases over time. Importantly, DeepGaze II, a model which includes high-level features, has the largest advantage at the beginning of this phase, i.e., the advantage of including high-level features starts immediately and reaches its peak already at fixation \#2. As all predictions decline in parallel, a reason for the decline might be an increase of fixations which are not guided by the scene at all.

Finally, in the last phase, the fixation density reaches an equilibrium and all fixation numbers predict each other equally well. Although subjects preferentially look at the same locations they look at during the main exploration, they are overall less predictable.  

In the search data we find a qualitatively similar temporal evolution of the fixation density as for memorisation. We again see an onset response with initial central fixation bias, a period of marginally better predictability and a final equilibrium state. However, the fixation density is much less predictable in general, there is virtually no central fixation bias after the onset response and all saliency models perform much worse in predicting fixation locations, especially when we reuse the mapping from saliency map to fixation density from the scene viewing dataset. 
The initial central fixation bias is weaker in this dataset as we delayed the onset of the first saccade \citep{rothkegel2017}. 

\subsection{Low-level vs.~high-level}

At first glance, the observation that low-level models predict fixations well at the beginning and worse later in the trial fits well with the classical saliency model idea that the initial exploration is driven by low-level bottom-up factors. However, the performance decline of low-level models resembles the decline of the empirical density, early fixations are well predicted by later fixation densities and adding high-level features as in DeepGaze II improves predictions throughout the trial. These findings rather suggest that during the initial main exploration fixations are driven by the same high-level features as later fixations.  

Indeed, even within the first fixation, adding high-level information improves predictions. Starting $200ms$ after image onset DeepGaze II performs better than the early vision based model. Nonetheless, low-level models perform best for the first fixation and have the largest advantage over the central fixation bias for the first fixation. This increase in early vision based model performance for the first fixation corroborates earlier findings, that low-level guidance influences mainly the first fixation \citep{anderson2015,anderson2016}.

After the initial onset response our data are even compatible with the extreme stance that low-level features have no influence on eye movement behaviour. This account agrees well with a range of literature which shows influences of objects \citep{einhauser2008b, stoll2015} and other high-level features \citep{henderson1999b,torralba2006} on eye movements. The predictive value of low-level features, like contrast at a location, could then be explained by their correlation with being interesting in a high-level sense. Such correlations are expected, because very low contrast areas are devoid of content. As such, this explanation would also work to explain high-level influences based on low-level features. However, high-level features are better at predicting, such that they necessarily have some predictive value beyond low-level features. Also, manipulations of contrast seem to have little influence on the fixation distribution beyond the first fixation \citep{acik2009, anderson2015}, such that the part of the fixation distribution, which could be explained both by low-level and by high-level features, is more likely to be explained by high-level features.

Adjusting the early vision based model to the target subjects search for barely improves model performance. This finding implies that merely reweighing low-level features is insufficient for modelling eye movements in visual search. This failure argues against models in which simple top-down control operates on low-level features to guide eye movements \citep{itti2000b, treisman1980, wolfe1994}. More complex processing of low-level features resulting for example in optimal control during visual search \citep{najemnik2008} is compatible with our data. How well a specific target could be detected by the observer at a location is not a low-level feature or a simple combination of low-level features any more. Thus, we do not consider optimal control in search a low-level feature theory.

Based on these considerations, low-level features seem relatively unimportant for eye movement control in natural scenes and are largely
restricted to an early bottom-up response during a trial. One reason to explain this lack of effect might be that we used stationary scenes. Instead, onsets or movements might be necessary to attract fixations against top-down control \citep{jonides1988,yantis1990}. 
Moving scenes can produce much higher coherence among eye movements of participants \citep{dorr2010} and the classic experiments showing bottom-up control all used sudden onsets \citep[for example]{hallett1978}. 

A normative reason why eye movement control should focus on high-level features in stationary stimuli might be that fast responses are only required if the stimulus changes. When the stimulus changes a fast response based on simple features might be advantageous, but if the stimulus is stationary the eye movement control system has sufficient time for more complex computations.

\subsection{Bottom-up vs.~top-down}
Based on the search results, we can confirm earlier reports that fixation locations during visual search are hardly predicted by saliency models \citep{chen2006, henderson2007, einhauser2008a}, which shows that top-down control can overwrite bottom-up control when subjects view static natural scenes. We even see some influence of the target in our visual search data, which argues for a fairly detailed adjustment of eye movements to the concrete task at hand. This result fits well with earlier observations we made on this dataset \citep{rothkegel2018}, which  showed that subjects adjusted their saccade lengths and fixation durations to the target they searched for. Thus, our overall observations argue for a strong, detailed top-down influence on eye movement control during visual search.

This explanation implies that bottom-up factors can usually be overruled by top-down control signals as present in visual search.
The only exception to this argument might be the first fixation chosen by the observer, as the first chosen fixation  follows a different density than later fixations, is best predicted by saliency models and can even be predicted reasonably well in the visual search condition. 
Complicating the analysis of the first chosen fixation, we observe a temporal evolution within the first fixation from bottom-up to top-down control. This transition has been reported before as earlier saccades are more strongly biased towards the image center \citep{rothkegel2017} and might be driven more by bottom-up features \citep{anderson2015,anderson2016}. The transition from bottom-up effects to more value driven saccades within a single fixation duration was also observed in single saccade tasks with artificial stimuli \citep{schutz2012}. Thus the transition from bottom-up to top-down control might occur early, most likely already within the first fixation.

\subsection{Future prospects}
As we observe that the fixation density changes over the course of a trial, a single fixation density seems to be an insufficient description of eye movement control. Instead we found that exploring the temporal dynamics of eye movement behaviour throughout a trial provides interesting insights into the control of eye movements. These dynamics have been studied earlier \citep[e.g.][]{tatler2008,over2007} already providing interesting insights.

Additionally, investigating systematic tendencies in eye movement behaviour \citep{tatler2008,tatler2009} could be informative. Such tendencies might include behaviours like "scanning" or other systematic ways to search for a target. How these tendencies arise and interact with the image content are upcoming challenges for eye movement research.

To combine these observations into a coherent theory, models of eye movement behaviour will have to evolve to incorporate predictions over time and with dependencies between fixations. So far, there are few models which produce dependencies between fixations \citep[see][for notable exceptions]{engbert2015, clarke2017, lemeur2015, tatler2017} and even those who do are rarely evaluated regarding their abilities to produce natural dynamics and generally do not handle a connection to the explored images. Here we only scratch the surface of the possibilities to check models more thoroughly using the dynamics of eye movements. We believe that we now have the statistical methods \citep{barthelme2013,schutt2017} and datasets to pursue this research direction further.

To facilitate the exploration of the dynamical aspects of eye movement behaviour, we share the data from our scene viewing experiment at XXX and the data from our search experiment at XXX. We hope that such shared large datasets may provide a strong basis for the exploration of the dynamics of eye movements.

\bibliography{output.bib}
\bibliographystyle{jovcite}

\end{document}